\documentclass[aoas,nameyear,seceqn,dvips]{arximspdf}
\usepackage{graphics}
\doi{10.1214/10-AOAS326}
\volume{4}
\issue{3}
\pubyear{2010}
\firstpage{1476}
\lastpage{1497}

\makeatletter
\renewcommand{\citep}[1]{[\citet{#1}]}
\makeatother

\begin{document}
\begin{frontmatter}

\title{Prediction-based classification for\\ longitudinal
biomarkers\thanksref{T1}}
\pdftitle{Prediction-based classification for longitudinal
biomarkers}
\runtitle{PBC for longitudinal biomarkers}

\thankstext{T1}{Supported by the National Institutes of
Health (R01-AI056983 to A. S. F., R01-AI51225 to L.~J.~M. and U01-AI051986
to L. J. M.), the Philadelphia Foundation and the Fund from the
Commonwealth Universal Research Enhancement Program, Pennsylvania
Department of Health.}

\begin{aug}
\author[A]{\fnms{Andrea S.} \snm{Foulkes}\ead[label=e1]{foulkes@schoolph.umass.edu}\corref{}},
\author[B]{\fnms{Livio} \snm{Azzoni}},
\author[C]{\fnms{Xiaohong} \snm{Li}},\\
\author[D]{\fnms{Margaret A.} \snm{Johnson}},
\author[D]{\fnms{Colette} \snm{Smith}},
\author[E]{\fnms{Karam} \snm{Mounzer}}\\
and
\author[B]{\fnms{Luis J.} \snm{Montaner}}
\runauthor{A. S. Foulkes et al.}
\affiliation{University of Massachusetts, The Wistar Institute,
University of Massachusetts,
Royal Free Hampstead NHS Trust, Royal Free and University College Medical
School, Philadelphia Field Initiating Group for HIV Trials (FIGHT)\\ and
The Wistar Institute}
\address[A]{A. S. Foulkes \\
Division of Biostatistics\\
University of Massachusetts\\
404 Arnold House\\
715 N. Pleasant Street\\
Amherst, Massachusetts 01003\\
USA\\
\printead{e1}} 
\address[B]{L. Azzoni\\
L. Montaner\\
The Wistar Institute\\
3601 Spruce Street\\
Philadelphia, Pennsylvania 19104-4268\\
USA}
\address[C]{X. Li\\
BG Medicine Inc.\\
610 Lincoln Street\\
Waltham, Massachusetts 02451\\
USA}
\address[D]{M. Johnson\\
Colette Smith\\
Royal Free and\\
\quad University College Medical School\hspace*{2pt}\\
Hampstead Campus\\
Rowland Hill Street\\
London NW3 2PF\\
United Kingdom}
\address[E]{K. Mounzer\\
Philadelphia FIGHT\\
1233 Locust Street 5th Floor\\
Philadelphia,
Pennsylvania 19107\\
USA}
\end{aug}

\received{\smonth{2} \syear{2009}}
\revised{\smonth{9} \syear{2009}}

%
\begin{abstract}
Assessment of circulating CD4 count change over time in HIV-infected
subjects on antiretroviral therapy (ART) is a central component of
disease monitoring. The increasing number of HIV-infected subjects
starting therapy and the limited capacity to support CD4 count testing
within resource-limited settings have fueled interest in identifying
correlates of CD4 count change such as total lymphocyte count, among
others. The application of modeling techniques will be essential to
this endeavor due to the typically nonlinear CD4 trajectory over time
and the multiple input variables necessary for capturing CD4
variability. We propose a prediction-based classification approach
that involves first stage modeling and subsequent classification based
on clinically meaningful thresholds. This approach draws on existing
analytical methods described in the receiver operating characteristic
curve literature while presenting an extension for handling a
continuous outcome. Application of this method to an independent test
sample results in greater than $98\%$ positive predictive value for CD4
count change. The prediction algorithm is derived based on a cohort
of $n=270$ HIV-1 infected individuals from the Royal Free Hospital,
London who were followed for up to three years from initiation of ART.
A test sample comprised of $n=72$ individuals from Philadelphia and
followed for a similar length of time is used for validation. Results
suggest that this approach may be a useful tool for prioritizing
limited laboratory resources for CD4 testing after subjects start
antiretroviral therapy.
\end{abstract}

%

\end{frontmatter}

\section{Introduction} \label{sec:intro}
Chronic HIV infection results in the progressive depletion of CD4$+$ T
lymphocytes from both lymphoid tissues and peripheral blood. Thus, the
monitoring of peripheral blood CD4 count is the standard used in
decision-making concerning initiation of antiretroviral therapy (ART),
as well as monitoring response to ART over time. In 2002 and again in
2006, the World Health Organization (WHO) proposed guidelines for
administration of ARTs in an effort to provide a clear public health
approach to utilization of these limited, yet very powerful drugs \citep
{who2006}. This series of recommendations includes routine collection
and monitoring of CD4 counts to inform decisions regarding both
initiation and switching of drug regimens. However, this report also
acknowledges that collection of repeated CD4 counts may not be feasible
in resource-limited settings due to the high costs associated with such
monitoring. In these instances, clinicians are advised to initiate
therapy in patients with asymptomatic\vspace*{1pt} HIV disease if total lymphocyte
count (TLC) falls below 1200 cells$/$mm$^3$.

In this manuscript we consider modeling strategies for using
alternative surrogate markers within an acute window (3 years)
post-initiation of therapy. Since publication of the WHO guidelines,
several reports have been published on the clinical utility of
alternative surrogate markers for monitoring post-therapy response and
specifically the correlation between these markers and CD4 count
[\citet{bagc2007}; \citet{biss2006}; \citet{ferr2004};
\citet{maha2004}; \citet{kamy2004}; \citet{badrwood2003}; \citet{bede2003};
\citet{spac2003}; \citet{kuma2002}]. These investigations
involve both cross-sectional and longitudinal data and implement a
variety of straightforward analytical methods. Typically,
cross-sectional comparisons between CD4 count and TLC as well as
longitudinal comparisons between the change in each of these variables
over a specified time period are performed using correlation analysis
[\citet{badrwood2003}; \citet{kamy2004}; \citet{kuma2002}; \citet{spac2003}]. A summary of analytic
strategies described for these settings, and their potential
limitations, is given in the discussion; notably, the scientific
findings of these reports are variable.

In this manuscript we describe a prediction-based classification (PBC)
framework for predicting biomarker trajectories based on a binary
decision rule. PBC was originally described in the setting of
classifying HIV genetic variants that capture variability in a
cross-sectional response to ART [\citeauthor{fou2002} (\citeyear{fou2002,fou2003})]. Within this
framework, we present two estimation procedures that both involve first
stage modeling using a generalized linear mixed effect model (GLMM). In
the first case, we dichotomize the biomarker a priori and use a
logit link function. In this case, our approach reduces simply to
fitting a logistic model coupled with a receiver operator
characteristic (ROC) curve analysis, which is commonly applied in
practice though it has not been described for this setting. The second
estimation approach we present is based on fitting a linear mixed
effects model to the observed CD4 count, as measured on a continuous
scale. This later approach may offer improved predictive performance
since it incorporates the full range of the continuous scale data. We
describe both approaches further in Section~\ref{sec:methods}.
Section~\ref{sec:example} then illustrates the method through
application to two cohorts of HIV-1 infected individuals followed for
three years after initiation of ART. Some simple extensions are
described in Section~\ref{sec:extensions} and finally we offer a
discussion of how the approaches complement existing methods in
Section~\ref{sec:discussion}.

\section{Methods} \label{sec:methods}

Monitoring patient level CD4 counts over time may involve consideration
of the observed counts at a given time point, the percent change in
counts across a given period of time or some other function of patient
level data. In general, interest lies in determining whether this
function of the data is above or below a threshold value. For example,
in monitoring absolute CD4 counts, thresholds of $200$ and $350$ are
considered within well-established treatment administration guidelines.
A threshold of $20\%$, on the other hand, is common for monitoring the
percent change in CD4 between visits over time. We begin in this
section by describing a general modeling framework. We then present an
approach for predicting whether absolute CD4 is above a clinically
meaningful threshold, at each of multiple discrete time points. In
Section~\ref{sec:extensions} we consider extensions of this framework
that allow us to consider functions of the biomarker under study, such
as percentage change over a given time period.

\subsection{Generalized linear mixed effects model} \label{sec:glmm}

Consider the generalized linear mixed effects model (GLMM) given by
\begin{equation} \label{eq:mod}
g(E[\mathbf{Y}_i])=\mathbf{X}_i \beta+ \mathbf{Z}_i b_i,
\end{equation}
where $\mathbf{Y}_i=(Y_{i1},\ldots,Y_{in_i})^T$ is a vector of the
$n_i$ responses for individual $i$, $g(\cdot)$ is a link function,
$\mathbf{X}_i$ is the $n_i \times M$ corresponding design matrix across
$M$ covariates, $\beta$ is the fixed-effects parameter vector and $b_i
\stackrel{\mathrm{i.i.d.}}{\sim} \operatorname{MVN}(0,\mathbf{D})$. Here $\mathbf{Z}_i$ is the
design matrix for the random effects and will typically include both an
intercept and time component. One choice of $\mathbf{X}_i$ and
$\mathbf{Z}_i$ is offered in the example of Section~\ref{sec:example}
and includes time varying values of white blood cell count and
lymphocyte percentage. This model is a natural choice for this setting
since repeated measures are taken over time on the same individual and
the time points are unevenly spaced across individuals
\citep{fitzlairware2004}.

In this manuscript we consider two approaches to fitting the model of
equation~(\ref{eq:mod}). Since ultimately we are interested in predicting
whether CD4 count is above (or below) a given threshold, we begin by
modeling a dichotomized version of the observed CD4 data. We use the
notation $Y_{ij}^+$ to indicate this binary representation of the
observed data. That is, we define the dependent variable
$Y_{ij}^+=I(\mathrm{CD4}_{ij}>K)$, where $\mathrm{CD4}_{ij}$ is the CD4
count at the $j$th time point for individual $i$ and $K$ is set equal
to a clinically meaningful threshold. In this case, the canonical logit
link is used to model the resulting binary outcome. Formally, if we let
${\theta}_{ij}=E[Y_{ij}^+]=\Pr(Y_{ij}^+=1)$, then equation~(\ref{eq:mod})
reduces in this setting to
\begin{equation} \label{eq:logit}
\theta_{ij}=\frac{\exp[\mathbf{x}_{ij} \beta+ \mathbf{z}_{ij}
b_i]}{1+\exp[\mathbf{x}_{ij} \beta+ \mathbf{z}_{ij} b_i]},
\end{equation}
where $\mathbf{x}_{ij}$ and $\mathbf{z}_{ij}$ are the rows of $\mathbf{X}_i$ and $\mathbf{Z}_i$ respectively, corresponding to the $j$th
measurement for individual $i$.

Second, we explore the utility of using the full range of the CD4 count
data by modeling CD4 as a continuous variable. That is, we let
$Y_{ij}=\mathrm{CD4}_{ij}$ and $g(\cdot)$ be the identity function, so that
the model of equation~(\ref{eq:mod}) reduces to the linear mixed effects
model (LMM), given by
\begin{equation} \label{eq:linear}
Y_{ij}=\mathbf{x}_{ij} \beta+ \mathbf{z}_{ij} b_i + \varepsilon_{ij},
\end{equation}
where $\varepsilon_{ij} \sim N (0,\sigma^2)$ and $b_i \perp
\varepsilon_{ij}$. Since we ultimately aim to predict whether CD4 is above
a given threshold, we then derive a prediction rule based on the
estimated mean and variance components from this model.

\subsection{Prediction-based classification} \label{sec:algorithm}
In fitting the mixed effects model of equation~(\ref{eq:mod}), we use the
complete vector of observed data, given by $\mathbf{y}_i = (y_{i0}, \ldots
,y_{in_i})$, for all individuals in our learning sample. In general, we
want to make predictions for new individuals under the assumption that
\emph{only} baseline values of $y_i$, given by $y_{i0}$, are observed.
In the usual model fitting context, the predicted $y$ is generated
using the empirical Bayes estimates of $b_i$, given by $\widehat
{b}_i=E[b | \mathbf{y}_i]$. Notably, this conditions on this complete data
vector and thus is not applicable to our setting, in which only the
$y_{i0}$ are available. Thus, we need to arrive at an alternative
estimate of the random effects that conditions only on the observed
data for new individuals. We consider two approaches in the context of
the linear mixed model. In the first case, we replace $\mathbf{y}_i$ with
$\mathbf{X}_i \widehat{\beta}$ in the formula for $\widehat{b}_i$. This is
our primary approach, described in Section~\ref{sec:cont} and applied
in the example of Section~\ref{sec:example}. The second alternative we
consider is to replace $\mathbf{y}_i$ with the baseline measure $y_{i0}$,
which is presented as an extension in Section \ref{sec:extensions}.

\subsubsection{Binary outcome}

After fitting the model of equation~(\ref{eq:mod}), mean and variance
parameter estimates can be used to arrive at a predicted mean response
for individual $i$ at the $j$th time point. Consider first the case in
which we dichotomize CD4 count and fit the GLMM with a logit link, as
described by equation~(\ref{eq:logit}). In this case, we have the
predicted probability of CD4 count being above the threshold $K$ at the
$j$th time point for individual $i$ given by
\begin{equation} \label{eq:thetapred}
\widehat{\theta}_{ij}=\frac{\exp[\mathbf{x}_{ij} \widehat{\beta} +
\mathbf{z}_{ij} \widehat{b}_i]}{1+\exp[\mathbf{x}_{ij} \widehat{\beta}
+ \mathbf{z}_{ij} \widehat{b}_i]},
\end{equation}
where $\widehat{\beta}$ is a maximum likelihood estimate of
$\beta$ and $\widehat{b}_{i}=E[b_i | \mathbf{y}_i^+]$ is the conditional
mean of the random effects for individual $i$, given the observed data
$\mathbf{y}_i^+$. Numerical integration techniques, such as Gaussian
quadrature, are required for model fitting in this setting since no
simple, closed-form solutions to maximum likelihood estimation are available.

\begin{table}[b]
\tablewidth=204pt
\caption{Contingency table notation for a given $\alpha$-prediction rule}\label{tab:contingency}
\begin{tabular*}{204pt}{@{\extracolsep{\fill}}lcccc@{}}
\hline
& & \multicolumn{2}{c}{$\bolds{y_{ij}^+}$} & \\[-5pt]
&&\multicolumn{2}{c}{\hrulefill}\\
& & $\bolds{1}$ & $\bolds{0}$ & \textbf{Total} \\
\hline
$\widehat{y}_{ij,\alpha}^+$ & 1 & $n_{11}$ & $n_{12} $
& $n_{1\cdot}$ \\
& 0 & $n_{21}$& $n_{22}$ & $n_{2\cdot}$\\ \hline
& Total & $n_{\cdot1}$ & $n_{\cdot2}$ & $n_{\cdot\cdot}$ \\
\hline
\end{tabular*}
\end{table}

A simple approach to prediction in this case is to let the predicted
outcome, given by $\widehat{y}_{ij}$, equal $1$ if $\widehat{\theta
}_{ij} \ge0.50$ and $0$ otherwise, where $\widehat{\theta}_{ij}$ is
defined by equation~(\ref{eq:thetapred}). Alternatively, we may want to
choose a prediction rule that controls a clinically meaningful
attribute. For example, in the CD4 prediction setting, we may want to
control the false positive rate, defined as the proportion of
individuals predicted to be above a safety threshold, when in fact
their CD4 counts are below this safe limit. In this case, we define
multiple rules, termed $\alpha$-prediction rules, that are given by
\begin{equation} \label{eq:predrule}
\widehat{y}_{ij,\alpha}^+ =
\cases{ 1, &\quad if ${\theta}_{ij} \ge1-\alpha$,\cr
0, &\quad otherwise,
}
\end{equation}
where the unobserved $\theta_{ij}$ is replaced with the
estimate $\widehat{\theta}_{ij}$. Notably, in making predictions for
new individuals, the complete vector\vspace*{1pt} $\mathbf{y}^+$ is not available and,
thus, $\widehat{b}_{i}=E[b_i | \mathbf{y}_i^+]$ in
equation~(\ref{eq:thetapred}) cannot be calculated. In the example provided below, we
let $\widehat{b}_i=E[b_i]=0$ for all $i$ in our test sample. An
alternative approach for the linear model setting is described in
Section~\ref{sec:extensions}.

Based on a given $\alpha$-prediction rule, we can generate the
contingency table given in Table~\ref{tab:contingency}. Here the
$n_{kl}$'s are the corresponding cell counts for $k,l=1,2$. For
example, $n_{11}$ is the number of observations that are observed to be
above the threshold ($y_{ij}^+=1$) and predicted to be above the
threshold ($\widehat{y}_{ij,\alpha}^+=1$).\vspace*{1pt} The sensitivity of this rule
is defined as the probability of correctly predicting an observation as
being above the threshold among those responses that are in fact above
the threshold and is given algebraically as $\Pr(\widehat
{y}_{ij,\alpha}^+=1 | y_{ij}^+=1)=n_{11}/n_{\cdot1}$. The corresponding
specificity is given by $\Pr(\widehat{y}_{ij,\alpha}^+=0
|y_{ij}^+=0)=n_{22}/n_{\cdot2}$ and\vspace*{1pt} the false positive rate is $\mathrm{FP}_{\alpha}=1-\mathrm{specificity}=n_{12}/n_{\cdot2}$. Positive predictive
value (PPV) and negative predictive value (NPV) are given by $
(n_{11}/n_{1\cdot} )$ and $ (n_{22}/n_{2\cdot} )$, respectively.
By varying the value of $\alpha$ in equation~(\ref{eq:predrule}), we
generate multiple prediction rules and can construct a corresponding
receiver operator characteristic (ROC) curve, which offers a visual
representation of the trade-off between sensitivity and specificity.
Specifically, an ROC curve is defined as a plot of the false positive
rate ($x$-axis) and corresponding sensitivity ($y$-axis) for each of
multiple classifiers, in our case prediction rules. In our setting,
each $\alpha$-rule contributes one point to the ROC curve. We define
the optimal rule as the one that controls the FP rate at a specified
level, though alternative criterion are equally applicable.

Since the prediction rule given by equation~(\ref{eq:predrule}) depends
on an estimate of $\theta_{ij}$ that is derived based on the data, a
cross-validation approach is necessary to obtain accurate estimates of
predictive performance, including sensitivity and false positive rate.
The motivation for this stems from the need to characterize the ability
to make predictions on observations that did not contribute to the
model fitting procedure. In this manuscript, we use an independent test
sample to evaluate model performance. The approach proceeds as follows:
First, model parameters are estimated using data arising from what we
refer to as the learning sample. Second, the best $\alpha$-rule is
identified based on the trade-off between sensitivity and specificity,
again using the learning sample data. The estimates of predictive
performance (e.g., false positive rate) based on the learning sample
are referred to as resubstitution estimates as the data used for
estimating error rates are the same as those used for deriving the
prediction rule. Finally, measures of predictive performance for the
chosen $\alpha$-rule are reported based on applying the rule to an
independent data set, which we refer to as the test sample data. These
test sample estimates are considered unbiased reflections of predictive
performance, as independent data sets are used to generate the rule and
describe its performance.

\subsubsection{Continuous outcome} \label{sec:cont}

The prediction approach just described for a binary outcome involves
simply fitting a logistic regression model and then generating an ROC
curve based on several probability cutoffs. While, to our knowledge,
this has not been applied to the setting of modeling biomarker
trajectories over time and specifically to CD4 monitoring, similar
approaches are used in practice in other settings [\citet{tostwein1994}; \citet{tostbegg1988}]. One reason that this approach may not be
optimal for the present setting is that CD4 count is measured on a
continuous scale. We thus consider a simple extension of this approach
that takes into consideration the full range of the observed CD4 count
data. We begin by modeling $y_{ij}=\mathrm{CD4}_{ij}$ as a quantitative
biomarker, using the linear mixed effects model of equation~(\ref
{eq:linear}), and then derive a prediction approach similar to the one
described by equation~(\ref{eq:predrule}).

The model derived predicted value of $y_{ij}$ is given by $\widehat
{y}_{ij}^*=\mathbf{x}_{ij}\widehat{\beta} + \mathbf{z}_{ij}
\widehat{b}_i$. Here $\mathbf{x}_{ij}$ and $\mathbf{z}_{ij}$ are again
respectively the rows of $\mathbf{X}_{i}$ and $\mathbf{Z}_i$
corresponding to the $j$th measurement for individual $i$,
$\widehat{\beta}=\sum_{i=1}^N(\mathbf{X}_i^T\widehat
{\Sigma}_i^{-1}\mathbf{X}_i)^{-1}\mathbf{X}_i^T\widehat{\Sigma}_i\mathbf{y}_i$
is the least squares estimate of $\beta$, $\widehat{b}_i=E(b_i |
\mathbf{y}_i)=\widehat{\mathbf{D}}\mathbf{Z}_i^{T}\widehat{\Sigma}_i^{-1}(\mathbf{y}_i-\mathbf{X}_i\widehat{\beta})$
is the best linear unbiased predictor (BLUP) of the random effects for
individual $i$, $\widehat{\Sigma}_i =
\widehat{\operatorname{Var}}(\mathbf{y}_i)=\mathbf{Z}_i\widehat{\mathbf{D}}\mathbf{Z}_i^T
+ \widehat {\sigma}^2I$, and $\widehat{\mathbf{D}}$ and
$\widehat{\sigma}^2$ are the restricted maximum likelihood estimates of
$\mathbf{D}$ and $\sigma^2$, respectively. Rather than estimate
$\theta_{ij}=\Pr(\mathrm{CD4}_{ij}>K)$ of equation~(\ref{eq:predrule}),
we describe a one-sided prediction interval approach to identify a rule
that is similar to the one described by this equation.

First note that the lower bound of the one-sided $(1-\alpha)$
prediction interval for $y_{ij}$ is given by
\begin{equation} \label{eq:lowerbound}
l_{ij,\alpha} = \widehat{y}_{ij}- z_{\alpha} \sqrt{\operatorname{Var}  ( \widehat
{y}_{ij} - y_{ij}  )},
\end{equation}
where $z_{\alpha}$ is the quantile of a standard normal
corresponding to a $1-\alpha$ probability and $\operatorname{Var}  ( \widehat
{y}_{ij} - y_{ij}  )$ is referred to as the prediction variance.
In this manuscript, we treat this interval as an approximate credible
interval, so that we are $(1-\alpha)\%$ certain that the random
variable $Y_{ij}$ will be greater than this realization of the lower
bound. In other words, $\Pr  (Y_{ij}>l_{ij,\alpha}  )= (1-\alpha
)\%$. Thus, if $l_{ij,\alpha}>K$, we are at least $(1-\alpha)\%$
certain that $Y_{ij}>K$. In other words, $l_{ij,\alpha}>K$ is
equivalent to $\theta\ge(1-\alpha)$. As a result, the rule given by
%
\begin{equation} \label{eq:predrule2}
\widehat{y}_{ij,\alpha}^+=
\cases{ 1, & \quad if $l_{ij,\alpha} > K$, \cr
0, &\quad otherwise,
}
\end{equation}
is equivalent to the one given by equation~(\ref{eq:predrule}). As
described in \citet{mclesand1991} and \citet{McCuSear2001},
the prediction variance is given by
$\operatorname{Var}(\widehat{y}_{ij}-\mathbf{x}_{ij}\beta
-\mathbf{z}_{ij}b_i)
=\mathbf{x}_{ij}\operatorname{Var}(\widehat{\beta})\mathbf{x}_{ij}^T +
\mathbf{z}_{ij} \operatorname{Var}(\widehat{b}_i-b_i) \mathbf{z}_{ij}^T
+
\mathbf{x}_{ij}\operatorname{Cov}(\widehat{\beta},\widehat{b}_i-b_i)\mathbf{z}_{ij}^T$
where $\operatorname{Var}(\widehat{\beta}) = \sum_{i=1}^N (\mathbf{X}_i
\Sigma_i^{-1} \mathbf{X}_i^T)^{-1}$,
$\operatorname{Var}(\widehat{b}_i-b_i) = ( \frac{1}{\sigma^2}
\mathbf{Z}_i^T \mathbf{Z}_i+\mathbf{D}^{-1} )^{-1}$
$-\operatorname{Cov}(\widehat{\beta },\widehat{b}_i-b_i) \mathbf{X}_i^T
\Sigma_i^{-1}\mathbf{Z}_i\mathbf{D}$ and
$\operatorname{Cov}(\widehat{\beta},\widehat{b}_i-b_i)
=-\mathbf{D}\mathbf{Z}_i^T\Sigma _i^{-1}\mathbf{X}_i
\operatorname{Var}(\widehat{\beta})$. In our setting, we are interested
in the prediction variance for a new \emph{observed} value and thus
have an additional $\sigma^2$ term. That is, $\operatorname{Var}  (
\widehat{y}_{ij} - y_{ij}  )$ of equation~(\ref{eq:lowerbound}) is equal
to
$\operatorname{Var}(\widehat{y}_{ij}-\mathbf{x}_{ij}\beta-\mathbf{z}_{ij}b_i)
+ \sigma^2$. The appropriateness of treating the above prediction
interval as a credible interval depends on prior assumptions about the
parameters of our model. Since we are using this as a means of
generating a prediction rule, and not as a tool for inference, this
approximation seems reasonable. It also performs well in the example
provided in Section~\ref{sec:example}. A study of the relative
advantages of applying a fully Bayesian approach to approximating the
posterior predictive distribution for this data setting is ongoing
research.

Again a test sample is used to characterize model performance. In the
linear mixed modeling setting, we note that $\operatorname{Var}(\widehat{\beta})$,
$\widehat{\mathbf{D}}$ and $\widehat{\sigma}$ are estimated based on the
model fitting procedure that uses the learning sample data.\vspace*{1pt} The
remaining variance terms, $\operatorname{Var}(\widehat{b}_i-b_i)$ and $\operatorname{Cov}(\widehat
{\beta},\widehat{b}_i-b_i)$ as well as the design elements $\mathbf{x}_{ij}$ and $\mathbf{z}_{ij}$ used in the calculation of $l_{ij,\alpha}$
of equation~(\ref{eq:lowerbound}) are based on the test sample data.
Notably, in both modeling frameworks, the BLUPs of the random effects
can not be calculated for a new individual for whom the response $\mathbf{y}_{i}$ is not observed. One approach to handling this unobserved data
is to replace $\mathbf{y}_{i}$ in the formula for $\widehat{b}_i$ with
$\mathbf{X}_i\widehat{\beta}$ so that $\widehat{b}_i=\widehat{\mathbf{D}}\mathbf{Z}_i^{T}\widehat{\Sigma}_i^{-1}(\mathbf{X}_i\widehat{\beta}-\mathbf{X}_i\widehat{\beta})=0$. This results in reducing $\widehat{y}_{ij}$ to
$\widehat{y}_{ij}=\mathbf{x}_{ij}\widehat{\beta}$ and is consistent with
assigning each individual the estimated population average. In the
example below, we use the prediction variance from the usual regression
setting of $\operatorname{Var}(\widehat{y}_{ij}-y_{ij})= \mathbf{x}_{ij}\operatorname{Var}(\widehat{\beta
})\mathbf{x}_{ij}^T +{\sigma}^2$. This prediction variance is less than
the one described above; however, as we are varying $z_{\alpha}$ of
equation~(\ref{eq:lowerbound}) to generate a series of classification
rules, the magnitude of the interval is less relevant. An alternative
approach for handling the random effects in the linear mixed modeling
framework is described in Section~\ref{sec:extensions}.

\section{Example} \label{sec:example}

The approach described in Section~\ref{sec:methods} is applied to a
cohort of $N=270$ individuals from the Royal Free Hospital, London who
were followed for up to three years after initiation of ART. Detailed
information on the patient population and laboratory methods can be
found in \citeauthor{smit2003} (\citeyear{smit2003,smit2004}). The aim of our analysis is to
determine the utility of baseline CD4 count and repeated measures on
WBC and lymphocyte percentage for predicting CD4 counts over time. Our
approach uses the complete CD4 count data (across all time points) from
a learning sample to generate a model; predictions based on this model
are then made, for the resubstituted data as well as for an independent
test sample, assuming that we only observe the baseline values of CD4.
Consideration is given to two clinically meaningful CD4 count
thresholds: $K=200$ and $K=350$ cells$/$mm$^3$. All analyses are
performed using R version 2.7.1. The median length of follow-up is $25$
months and the interquartile range (IQR) for length of follow-up is
$(14,32)$ months. The median number of follow-up time points is $9$
with a full range of $2$ to $24$. In total, there are $2635$ records
including baseline measurements. The median baseline CD4 count for this
cohort is $219.5$ with an IQR equal to $(114, 333)$.

Linear and generalized linear mixed effects models are fitted in R
using the \texttt{lme()} and \texttt{lmer()} functions of the \texttt
{nlme} and \texttt{lme4} packages, respectively. We assume a piecewise
linear mixed effects model for modeling CD4 count after initiation of
ART \citep{fitzlairware2004}. This model is appropriate since CD4 count
tends to rise rapidly for approximately one month and then proceeds to
increase more gradually. Fixed effects for baseline CD4 count (on a log
base 10 scale), baseline and time varying values of WBC and lymphocyte
percentage and time before and after one month of follow-up are
included in the model as predictors. In addition, interactions between
each time component and baseline values of WBC and lymphocyte percent
are included.

The design matrix $\mathbf{X}_i$ for the fixed effects of equation~(\ref
{eq:mod}) is thus given by
\begin{eqnarray} \label{eq:design}
\mathbf{X}_i =  [ \matrix{1_N & \mathbf{X}_{i1} & \mathbf{X}_{i2} } ],
\nonumber\\[18pt]\\[-36pt]
\eqntext{\mathbf{X}_{i1} =  \left[
\matrix{ y_{i0} & w_{i0} & l_{i0} & t_{i1} & (t_{i1}-1)_+
& w_{i1} & l_{i1}
\cr
y_{i0} & w_{i0} & l_{i0} & t_{i2} & (t_{i2}-1)_+ & w_{i2} & l_{i2}
\cr
& && \vdots&& & \cr
y_{i0} & w_{i0} & l_{i0} & t_{in_i} & (t_{in_i}-1)_+ & w_{in_i} &
l_{in_i}
}
\right],\hspace*{49pt}}
\\
\eqntext{\mathbf{X}_{i2} =  \left[
\matrix{
t_{i1}*w_{i0} & t_{i1}*l_{i0} & (t_{i1}-1)_+*w_{i0} &
(t_{i1}-1)_+*l_{i0} \cr
t_{i2}*w_{i0} & t_{i2}*l_{i0} & (t_{i2}-1)_+*w_{i0} &
(t_{i2}-1)_+*l_{i0} \cr
& \vdots& & \cr
t_{in_i}*w_{i0} & t_{in_i}*l_{i0} & (t_{in_i}-1)_+*w_{i0}&
(t_{in_i}-1)_+*l_{i0}
}
\right],}
\end{eqnarray}
where $w_{i0}$ and $l_{i0}$ are respectively baseline WBC and
baseline lymphocyte percent, $t_{ij}$ is time in months since
initiation of ART, $(t_{ij}-1)_+$ is follow-up time after the first 1
month on ART for $t_{ij}>1$ and $0$ otherwise, and $w_{ij}$ and
$l_{ij}$ are respectively WBC and lymphocyte percent at time $t_{ij}$.
We define $y_{i0}$ in $\mathbf{X}_{i1}$ as $\log(\mathrm{CD4})$ for both the
linear and generalized linear model although the response variable,
given by $\mathbf{Y}_i=(Y_{i1},\ldots, Y_{in_i})$, is dichotomized for the
generalized linear model setting. Notably, this model allows for two
linear time trends, before and after 1 month of follow-up on ART.
Random person specific intercepts and slopes before the knot are also
assumed so that the design matrix $\mathbf{Z}_i$ for the random effects of
equation~(\ref{eq:mod}) is given by
\begin{equation} \label{eq:design1}
\mathbf{Z}_i = \left[
\matrix{ 1 & t_{i1} \cr
1 & t_{i2} \cr
& \vdots\cr
1 & t_{in_i} }
 \right].
\end{equation}
The random effects vector in equation~(\ref{eq:mod}) is given
by $b_i^T=  [
\matrix{ b_{i0} & b_{i1}
}
 ]$ representing the intercept and slope before the change point
for individual $i$.

We begin by fitting the generalized linear model, as described in
equation~(\ref{eq:logit}). In this case, post-baseline CD4 counts are
dichotomized and used as the outcome in the model fitting procedure.
Predicted probabilities of being above the CD4 threshold are estimated
for each post-baseline time point for each individual. The results of
applying a probability cutoff of $0.50$ are given in Table~\ref
{tab:glmmcounts}(a). We call this the ``naive'' approach since the
cutoff does not incorporate information about the resulting prediction
rule. While the sensitivities of these predictions rules ($0.98$ and
$0.90$) are high for both thresholds, the corresponding false positive
rates are also high ($0.54$ and $0.28$). This approach thus may not be
appropriate for CD4 testing since it yields a high probability of
falsely predicting that an individual's CD4 count is within a safe limit.

\begin{table}[b]
\caption{Observed and predicted counts (based on learning sample data)}\vspace*{1pt}
\label{tab:glmmcounts}
(a) GLMM approach with a ``naive'' 0.50 probability
cutoff
\begin{tabular*}{215pt}{@{\extracolsep{\fill}}lcccc@{}}
\hline
& & \multicolumn{2}{c}{\textbf{Observed}} & \\[-5pt]
&&\multicolumn{2}{c}{\hrulefill}\\
& & $\bolds{{>}200}$ & $\bolds{{<}200}$ & \textbf{Total} \\
\hline
Predicted & $>$200 & $1932$ & $215$ & $2147$ \\
& $<$200 & $\hphantom{00}34$& $184$ & $218$ \\[3pt]
& Total & $1966$ & $399$ & $2365$ \\
\hline\\[-14pt]
\hline
& & \multicolumn{2}{c}{\textbf{Observed}} & \\[-5pt]
&&\multicolumn{2}{c}{\hrulefill}\\
& & $\bolds{{>}350}$ & $\bolds{{<}350}$ & \textbf{Total} \\
\hline
Predicted & $>$350 & $1194$ & $\hphantom{0}289$ & $1483$ \\
& $<$350 & $\hphantom{0}137$& $\hphantom{0}745$ & $882$ \\[3pt]
& Total & $1331$ & $1034$ & $2365$ \\
\hline
\end{tabular*}
\end{table}
\setcounter{table}{1}
\begin{table}
\caption{Continued}
\begin{tabular}{@{}lccccccccc@{}}
\multicolumn{10}{@{}c@{}}{(b) GLMM approach}\\
\hline
& & &\multicolumn{2}{c}{\textbf{Observed}} & && \multicolumn{2}{c}{\textbf{Observed}} & \\[-5pt]
&&&\multicolumn{2}{c}{\hrulefill}&&&\multicolumn{2}{c}{\hrulefill}\\
& && $\bolds{{>}200}$ & $\bolds{{<}200}$ & \textbf{Total} && $\bolds{{>}200}$ & $\bolds{{<}200}$ & \textbf{Total} \\
\hline
Predicted\tabnoteref{tabl1} & $>$200 && $\hphantom{0}826$ & $\hphantom{0}18$ & $\hphantom{0}844$ && 1206 & \hphantom{0}37
& 1243 \\
& $<$200 && $1140$& $381$ & $1521$ && \hphantom{0}760 & 362 & 1122 \\[3pt]
& Total && $1966$ & $399$ & $2365$ && 1966 & 399 & 2365 \\
\hline\\[-14pt]
\hline
& && \multicolumn{2}{c}{\textbf{Observed}} & && \multicolumn{2}{c}{\textbf{Observed}} & \\[-5pt]
&&&\multicolumn{2}{c}{\hrulefill}&&&\multicolumn{2}{c}{\hrulefill}\\
& && $\bolds{{>}350}$ & $\bolds{{<}350}$ & \textbf{Total} & &$\bolds{{>}350}$ & $\bolds{{<}350}$ & \textbf{Total} \\
\hline
Predicted\tabnoteref{tabl1} & $>$350 && $\hphantom{0}669$ & $\hphantom{00}50$ & $\hphantom{0}719$ && \hphantom{0}880 & \hphantom{0}103
& 983 \\
& $<$350 && $\hphantom{0}662$& $\hphantom{0}984$ & $1646$ && \hphantom{0}451 & \hphantom{0}931 & 1382 \\[3pt]
& Total && $1331$ & $1034$ & $2365$ && 1331 & 1034 & 2365 \\
\hline
\end{tabular}\vspace*{5pt}
\begin{tabular}{@{}lccccccccc@{}}
\multicolumn{10}{@{}c@{}}{(c) LMM approach}\\
\hline
& && \multicolumn{2}{c}{\textbf{Observed}} & && \multicolumn{2}{c}{\textbf{Observed}} & \\[-5pt]
&&&\multicolumn{2}{c}{\hrulefill}&&&\multicolumn{2}{c}{\hrulefill}\\
&& & $\bolds{{>}200}$ & $\bolds{{<}200}$ & \textbf{Total} && $\bolds{{>}200}$ & $\bolds{{<}200}$ & \textbf{Total} \\
\hline
Predicted\tabnoteref{tabl1} & $>$200 && $1291$ & $\hphantom{0}19$ & $1319$ && 1558 &
\hphantom{0}38 & 1596 \\
& $<$200 && $\hphantom{0}675$& $380$ & $1055$ && \hphantom{0}408 & 361 & \hphantom{0}769 \\[3pt]
& Total && $1966$ & $399$ & $2365$ && 1966 & 399 & 2365 \\
\hline\\[-14pt]
\hline
& && \multicolumn{2}{c}{\textbf{Observed}} & && \multicolumn{2}{c}{\textbf{Observed}} & \\[-5pt]
&&&\multicolumn{2}{c}{\hrulefill}&&&\multicolumn{2}{c}{\hrulefill}\\
& && $\bolds{{>}350}$ & $\bolds{{<}350}$ & \textbf{Total} && $\bolds{{>}350}$ & $\bolds{{<}350}$ & \textbf{Total} \\
\hline
Predicted\tabnoteref{tabl1} & $>$350 && $\hphantom{0}760$ & $\hphantom{00}51$ & $\hphantom{0}811$ && \hphantom{0}940 & \hphantom{0}103
& 1043 \\
& $<$350 && $\hphantom{0}571$& $\hphantom{0}983$ & $1554$ && \hphantom{0}391 & \hphantom{0}931 & 1322 \\[3pt]
& Total && $1331$ & $1034$ & $2365$ && 1331 & 1034 & 2365 \\
\hline
\end{tabular}
\tabnotetext{tabl1}{Predicted counts are based on rules with resubstitution FP rate
estimates of approximately (but not greater than) $5\%$ (left panels)
and $10\%$ (right panels).}
\end{table}

\begin{figure}
\begin{tabular}{c}

\includegraphics{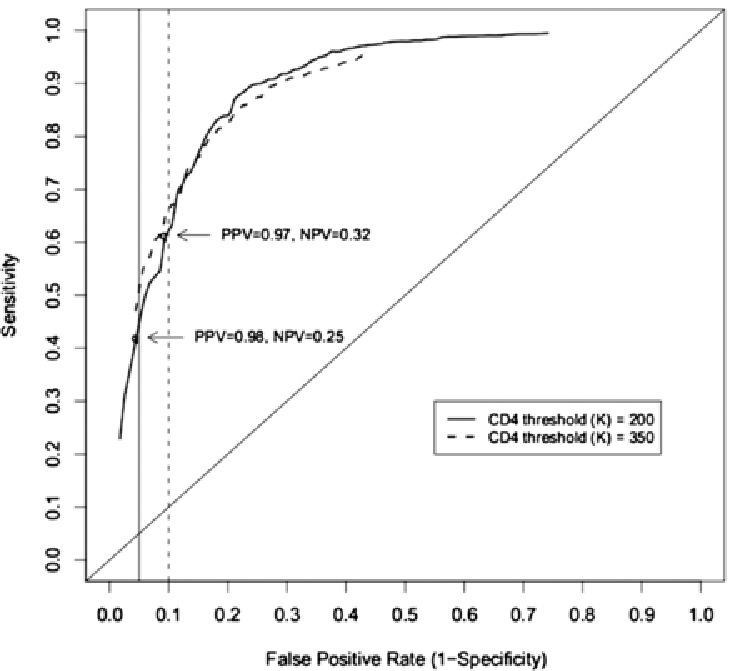}
\\
(a)
\\[6pt]

\includegraphics{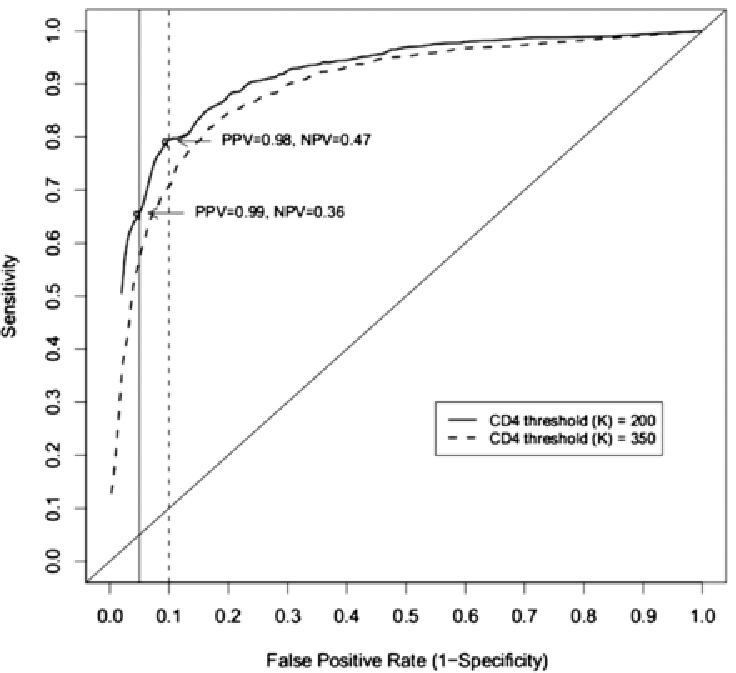}
\\
(b)
\end{tabular}
\caption{ROC curves based on resubstitution estimates. \textup{(a)} GLMM, \textup{(b)} LMM.} \label{fig:londonresub}
\end{figure}

Next, several $\alpha$ cutoffs are considered to generate multiple
prediction rules and an ROC curve is generated, as illustrated in
Figure~\ref{fig:londonresub}(a). This is again based on the GLMM
approach to model fitting. Data corresponding to rules with
resubstitution FP rates of approximately (but not greater than) $5\%$
and $10\%$ and CD4 threshold cutoffs of $K=200$ and $350$ are provided
in Tables~\ref{tab:glmmcounts}(b) and (c). Resubstitution-based summary
measures are given in Table~\ref{tab:lssum}(a). Based on a CD4
threshold of $K=200$, a FP rate of $0.09$ corresponds to a sensitivity
of $0.61$, a positive predictive value of $0.97$ and a negative
predictive value of $0.32$. For the same CD4 threshold, a FP of $0.05$
corresponds to a sensitivity of $0.42$, a positive predictive value of
$0.98$ and a negative predictive value of $0.25$.

\begin{table}
\tabcolsep=0pt
\caption{Estimates of predictive performance} \label{tab:lssum}
\begin{tabular*}{\textwidth}{@{\extracolsep{\fill}}lcccccccc@{}}
\multicolumn{9}{c}{(a) GLMM approach}\\
\hline
& \multicolumn{4}{c}{\textbf{GLMM (LS)}} & \multicolumn{4}{c@{}}{\textbf{GLMM (TS)}} \\[-5pt]
&\multicolumn{4}{c}{\hrulefill}&\multicolumn{4}{c@{}}{\hrulefill}\\
& \textbf{Sens} & \textbf{Spec} & \textbf{PPV} & \textbf{NPV} & \textbf{Sens} & \textbf{Spec} & \textbf{PPV} & \textbf{NPV} \\
\hline
$K=200$: & & & & & & & & \\
$\mathrm{LS\ FP}<0.05$ & 0.42 & 0.95 & 0.98 & 0.25 & 0.66 & 0.96 & 0.99 & 0.31 \\
$\mathrm{LS\ FP}<0.10$ & 0.61 & 0.91 & 0.97 & 0.32 & 0.77 & 0.96 & 0.99 & 0.39 \\[3pt]
$K=350$: & & & & & & & & \\
$\mathrm{LS\ FP}<0.05$ & 0.50 & 0.95 & 0.93 & 0.60 & 0.61 & 0.95 & 0.95 & 0.59 \\
$\mathrm{LS\ FP}<0.10$ & 0.66 & 0.90 & 0.90 & 0.67 & 0.79 & 0.90 & 0.93 & 0.71 \\
\hline
\end{tabular*}\\
\begin{tabular*}{\textwidth}{@{\extracolsep{\fill}}lcccccccc@{}}
\multicolumn{9}{c}{(b) LMM approach}\\
\hline
& \multicolumn{4}{c}{\textbf{LMM (LS)}} & \multicolumn{4}{c@{}}{\textbf{LMM (TS)}} \\[-5pt]
&\multicolumn{4}{c}{\hrulefill}&\multicolumn{4}{c@{}}{\hrulefill}\\
& \textbf{Sens} & \textbf{Spec} & \textbf{PPV} & \textbf{NPV} & \textbf{Sens} & \textbf{Spec} & \textbf{PPV} & \textbf{NPV} \\
\hline
$K=200$: & & & & & & & & \\
$\mathrm{LS\ FP}<0.05$ & 0.66 & 0.95 & 0.99 & 0.36
& 0.77 & 0.96 & 0.99 & 0.39\\
& (0.60, 0.75)&& (0.98, 0.99)&(0.31, 0.46) \\
$\mathrm{LS\ FP}<0.10$ & 0.79 & 0.90 & 0.98 & 0.47
& 0.84 & 0.96 & 0.99 & 0.49 \\
& (0.74, 0.83)&& (0.97, 0.98)&(0.37, 0.54) \\[3pt]
$K=350$: & & & & & & & & \\
$\mathrm{LS\ FP}<0.05$ & 0.57 & 0.95 & 0.94 & 0.63
& 0.73 & 0.93 & 0.95 & 0.67 \\
& (0.44, 0.67)&& (0.92, 0.95)&(0.56, 0.70) \\
$\mathrm{LS\ FP}<0.10$ & 0.71  & 0.90 & 0.90 & 0.70
& 0.84 & 0.90 & 0.94 & 0.77 \\
&(0.65, 0.79)&& (0.88, 0.92)&(0.66, 0.78) \\
\hline
\end{tabular*}
\end{table}

Next we fitted the linear mixed effects model, as described by
equation~(\ref{eq:linear}), to the observed CD4 count data. The resulting
ROC curve illustrating the sensitivity and corresponding false positive
rates in this cohort (resubstitution estimates) is given in Figure~\ref
{fig:londonresub}(b). Count data corresponding to rules for which
thresholds are $K=200$ and $350$ and the resubstitution FP rates are
approximately (but not greater than) $5\%$ and $10\%$ are given in
Table~\ref{tab:glmmcounts}(b). Corresponding summaries, as well as $95\%
$ bootstrap confidence intervals (CIs), are reported in Table~\ref
{tab:lssum}(b). To arrive at CIs, we repeatedly sample individuals with
replacement and in each case, fit a linear mixed effects model. The
prediction rule corresponding to FP rates of approximately (but not
greater than) $5\%$ and $10\%$ are selected and corresponding
resubstitution estimates of sensitivity, PPV and NPV are recorded. A
total of $100$ bootstraps are performed for each threshold and the
fifth and ninety-fifth percentiles reported.

Based on a CD4 cutoff of $200$, a FP rate of $0.10$ corresponds to a
sensitivity of $0.79$ [$95\%$ CI (0.74, 0.83)]. In this case, the PPV
is $0.98$ (0.97, 0.98) and the NPV is $0.47$ (0.37, 0.54). This
corresponds to the rule in which $\alpha=0.035$. That is, an
individual's CD4 count is predicted to be above 200 if the probability
that this measurement is greater than 200 is at least $1-0.035=96.5\%$.
For the same CD4 threshold, a FP rate of $0.05$ corresponds to a
sensitivity of $0.66$ (0.60, 0.75), PPV of $0.99$ (0.98, 0.99) and NPV
of $0.36$ (0.31, 0.46).

In order to further evaluate model performance, we apply our prediction
rule to $399$ observations across $n=72$ individuals from an
independent cohort in Philadelphia. We use only baseline CD4 counts to
make predictions, assuming that this is all that is available. The
median baseline CD4 in this cohort is $260.5$~cells$/$mm$^3$ and the IQR
is $(159.0, 354.2)$. Test sample estimates for sensitivity, false
positive rate, PPV and NPV are provided in Tables~\ref{tab:lssum}(a)
and (b) for each of the prediction rules. A tabular summary of counts
for one rule based on the LMM approach is given in Table~\ref
{tab:philadelphia}. The total count is $n= 327$ since there are
$399-72=327$ post-baseline measurements for this cohort. In this case,
$n=240$ measurements are predicted to be above the threshold, while
$87$ are predicted below. Since this is intended as a prioritization
tool, this rule would suggest performing a true CD4 test on the $87$
observations that are predicted below the threshold to confirm the true
value. A ``savings'' associated with this rule is $240/327=73\%$ since a
CD4 test would not be required for this percentage of the observations.
The ``cost'' is the associated false positive rate of $2/45=4.4\%$.
Interestingly, the test sample estimates based on the LMM approach
[Table~\ref{tab:lssum}(b)] appear slightly better than the
resubstitution estimates. In fact, in some cases, these test sample
estimates are greater than the $95\%$ bootstrap confidence limits
derived based on the learning sample. This result may be a consequence
of the overall slightly higher baseline CD4 count in the Philadelphia
(test sample) cohort. A discussion of the potential utility of
stratified analysis (e.g., according to baseline CD4 counts) is
provided in Section~\ref{sec:discussion}.

\begin{table}
\tablewidth=216pt
\caption{Observed and predicted counts (based on test sample data)}
\label{tab:philadelphia}
\begin{tabular*}{216pt}{@{\extracolsep{\fill}}lcccc@{}}
\hline
& & \multicolumn{2}{c}{\textbf{Observed}} & \\[-5pt]
&&\multicolumn{2}{c}{\hrulefill}\\
& & $\bolds{{>}200}$ & $\bolds{{<}200}$ & \textbf{Total} \\
\hline
Predicted & $>$200 & $238$ & $\hphantom{0}2$ & $240$ \\
& $<$200 & $\hphantom{0}44$& $43$ & $\hphantom{0}87$ \\[3pt]
& Total & $282$ & $45$ & $327$ \\ \hline
\end{tabular*}
\end{table}

\section{Extensions} \label{sec:extensions}
In this section we briefly describe two extensions of the method
outlined in Section~\ref{sec:methods} to illustrate its flexibility and
directions for further development. First, we consider one approach to
incorporating information about the individual level random effects
into our prediction algorithm for the linear mixed effects setting.
This approach is relevant as it provides a potential framework for
incorporating observed, post-baseline CD4 counts into the model.
Additionally, it illustrates the trade-off between using baseline data
within the fixed effects design matrix, and using these data to inform
prediction of the random effects. Second, we detail how this method can
be applied to making predictions about changes in CD4 count over time.
Extensions for modeling alternative outcomes are relevant, as clinical
decision making generally takes into account both absolute and relative
CD4 count changes.

\subsection{Using observed response data to inform BLUPs of random effects}

While leading to a prediction rule with good predictive performance,
the approach described in Section~\ref{sec:methods} does not take into
account the latent effects that result in some individuals having
higher or lower responses, information that is typically captured in
random effects. Several alternatives exist. For example, the prediction
variance used in the example above is based on the usual regression
setting, $\operatorname{Var}(\widehat{y}_{ij}-y_{ij})=\operatorname{Var}(x_{ij}\widehat{\beta
}-x_{ij}\beta-\varepsilon)$. Alternatively,\vspace*{1pt} we could use $\operatorname{Var}(\widehat
{y}_{ij}-y_{ij})=\operatorname{Var}(x_{ij}\widehat{\beta}-x_{ij}\beta
-z_{ij}b_i-\varepsilon)=\mathbf{x}_{ij}\operatorname{Var}(\widehat{\beta})\mathbf{x}_{ij}^T +
\mathbf{z}_{ij}\widehat{\mathbf{D}}\mathbf{z}_{ij}^T + {\sigma}^2$. That is, while
we let $\widehat{b}_i=0$, we still include the true $b_i$ in the
prediction variance formula. Based on the London data, this results in
slight, yet unremarkable improvements in sensitivity (results not shown).

We can also estimate the random effects for new individuals based on
baseline data. In the example provided, we assume only baseline CD4
counts are available, and these are used in the fixed effects design
matrix rather than informing the random effects. To begin, we propose
fitting the model of equation~(\ref{eq:linear}) with the slight
modification that the observed baseline CD4 count, given by $y_{i0}$,
is now included in the response vector $\mathbf{Y}_i$ and removed from
the design matrix $\mathbf{X}_i$. In order to estimate the random
effects for a new individual (whose complete response vector
$\mathbf{y}_{i}$ is unobserved), we calculate the conditional
expectation of the random effects, given the baseline (observed)
response $y_{i0}$. That is, we replace $\widehat{b}_i= E(b_i |
\mathbf{y}_{i})$ with $\tilde{b}_{i}= E(b_i | y_{i0}) =
(y_{i0}-\mathbf{x}_{i0}\widehat{\beta}_0) / (\widehat
{D}_{1,1}+\widehat{\sigma}_{\varepsilon}^2) \widehat{D}_{1,\cdot}$,
where $\widehat{D}_{1,1}$ is the $(1,1)$ element of
$\widehat{\mathbf{D}}$ corresponding to the estimated\vspace*{2pt} variance of the
intercept random effect, $\widehat{D}_{1,\cdot}$ is the column vector
corresponding to the first column of $\widehat{\mathbf{D}}$,
$\widehat{\beta}_0$ is the first element of $\widehat{\beta}$
corresponding to the intercept fixed effect and $\mathbf{x}_{i0}$ is
the first row of $\mathbf{X}_i$. This equation is derived simply by
replacing the matrix $\mathbf{Z}_i$ with its first row and replacing
the vectors $\mathbf{y}_i$ and $\mathbf{X}_i\widehat{\beta}$ with their\vspace*{1pt}
first elements in the formula $\widehat{b}_i=E(b_i |
\mathbf{y}_i)=\widehat{\mathbf{D}}\mathbf{Z}_i^{T}\widehat{\Sigma}_i^{-1}(\mathbf{y}_i-\mathbf{X}_i\widehat{\beta})$.

Notably, this is not the same prediction of $b_{i}$ that would have
been arrived at if the complete data vector $\mathbf{y}_i$ were observed
and so the alternative notation $\tilde{b}_{i}$ is used. Through use of
the first column of the $\widehat{\mathbf{D}}$ matrix, we draw on the estimated
covariance between the random effects to fill in values for both the
intercept and slope random effects for each individual, while only
relying on baseline values of the response. Finally, we additionally
replace $\operatorname{Var}(\widehat{b}_i-b_i)$ and $\operatorname{Cov}(\widehat{\beta},\widehat
{b}_i-b_i)$ with $\operatorname{Var}(\tilde{b}_i-b_i)$ and $\operatorname{Cov}(\widehat{\beta},\tilde
{b}_i-b_i)$, respectively in the formula for $\operatorname{Var}(\widehat
{y}_{ij}-y_{ij})$. Applications of this approach to the London data
(results not shown) are similar to those reported, suggesting that, in
this data example, using the modified BLUPs in place of treating
baseline as CD4 as a predictor variable does not improve our prediction
algorithm. Observed post-baseline measures of CD4 that occur prior to
the time of prediction could be incorporated similarly into the
predicted random effects.

\subsection{Making predictions about the percentage change in CD4 count
over time}

In Sections~\ref{sec:methods} and~\ref{sec:example} we focus on the
setting in which interests lie in predicting the response at a single
time point. More generally, we may want to make a prediction about a
function of the CD4 counts for individual $i$ across a combination of
time points $j$. For example, we may be interested in the percentage
change in CD4 count over a specified period of time, given by the
function $f_t(Y_{ij})=(Y_{ij}-Y_{ij'})/Y_{ij}$ where $(j-j')=t$. We can
again begin by fitting the linear model of equation~(\ref{eq:linear}) to
the repeated CD4 count measures and arriving at predictions for new
observations based on this model. The predicted percentage change for a
single individual $i$ is then given by $\widehat{f}_t(y_{ij})= (\widehat
{y}_{ij}-\widehat{y}_{ij'}) / \widehat{y}_{ij}$. In order to determine
the prediction variance of $\widehat{f}_t({y}_{ij})$, we use the
multivariate delta method. Based on a first-order Taylor series
expansion, we have $\operatorname{Var}  [ \widehat{f}_t({y}_{ij})  ] = \operatorname{Var}
 [ (\widehat{y}_{ij}-\widehat{y}_{ij'}) / \widehat{y}_{ij}  ]
= \operatorname{Var}  [ \widehat{y}_{ij'} / \widehat{y}_{ij}  ] \approx U^TV
U$, where $U^T=  \pmatrix{ 1/\widehat{y}_{ij} & -\widehat{y}_{ij'} / \widehat
{y}_{ij}^2}$ is the score vector and $V$ is the variance--covariance matrix
of $  \pmatrix{ \widehat{y}_{ij} & \widehat{y}_{ij'}}^T$. The matrix $V$ is calculated using the same formula as for
$\operatorname{Var}(\widehat{y}_{ij})$ above, where the vectors $\mathbf{x}_{ij}$ and
$\mathbf{z}_{ij}$ are replaced by matrices with rows corresponding to the
time points $j$ and $j'$. Further exploration of the utility of fitting
a LMM and identifying an associated prediction rule for the percentage
change in CD4 count, or a rule that evaluates simultaneously the
absolute level and the percentage change within the PBC framework, is
ongoing research.

\section{Discussion} \label{sec:discussion}
This manuscript presents an analytic approach, which we term PBC, for
predicting a quantitative biomarker trajectory over time that combines
the generalized linear mixed effects model with an ROC curve type
approach. Two approaches to approximating the prediction rule of
equation~(\ref{eq:predrule}) are considered. In the first case, we
dichotomize the data a priori and model the resulting binary
outcomes over time; a generalized linear mixed effects modeling
approach is applied for direct estimation of $\theta_{ij}$. Since we
ultimately aim to arrive at a binary prediction rule, this approach is
intuitively appealing and consistent with applications of the logistic
model for prediction. In the second case, we model the data using a
linear mixed effects model, a standard approach to the analysis of
unevenly spaced, repeated measures data with a continuous response and
multiple predictor variables. The results of this model fitting
procedure are used in turn to inform predictions, in this case using a
rule that involves the lower bound of the corresponding prediction
interval. This second approach also offers intuitive appeal since it
allows for use of all of the observed data to inform the model fit. A
similar approach as the one described herein can be applied for
modeling pathogenesis, though the additional population level
variability in CD4 counts in the absence of therapy may lead to lower
predictive performance.

PBC differs in two regards from methods currently employed in this
setting. First, we apply first-stage modeling that can incorporate the
full range of multiple continuous and categorical predictors, as well
as quantitative data on our outcome (CD4 count) to inform our analysis.
Estimated mean and variance components from this model fitting
procedure are subsequently used to define a rule for predicting whether
a function of the observed CD4 count (within and across time points) is
above or below a clinically meaningful threshold. Multiple patient
level characteristics can be incorporated, including observed baseline
CD4 count and time-varying values of the potentially predictive markers
as described in Section~\ref{sec:example}. The proposed approach is
different from previously described approaches for this setting since
modeling is performed using all of the available data and a prediction
rule is associated with the resulting model. One potential advantage is
that we are able to draw on the full range of both the predictor and
outcome data to inform our investigation while still providing a binary
decision rule for clinical decision making based on resulting
probability estimates.

A second difference is that PBC provides a framework for modeling CD4
count trajectories over time that is not limited to characterizing
changes between two time points. Specifically, we consider models with
a single knot at one month after initiation of ART to account for the
rapid increase in CD4 count that is typically observed and the
subsequently slower rise over time [\citet{lairware1982}; \citet{fitzlairware2004}]. The GLMM is applied with individual
level random intercept and slope terms in order to account for the
within person correlation inherent in repeated measures data. The use
of a mixed effects model for longitudinal CD4 data has been described
for monitoring response to therapy \citep{maha2004}; however, the aim
of that investigation differed in that the investigators applied the
mixed model to uncover the within and between person variability in TLC
for fixed changes in CD4 count. In our setting, the mixed model is used
as a tool within a predictive algorithm that allows for prediction
across a temporal trajectory.

Several manuscripts also report receiver operating characteristic (ROC)
curve analyses using information on TLC as well as other markers, such
as hemoglobin to predict CD4 count. To our knowledge, all such
investigations involve a first-stage dichotomization of the proposed
markers as well as the outcome CD4 count. For example, Spacek et al.
describe an approach involving cutoff points for TLC ($<$1200
cells$/$mm$^3$ and $>$2000 cells$/$mm$^3$) and/or hemoglobin ($>$12~g$/$dl)
\citep{spac2003}, while others propose dichotomizing TLC based on
whether the change over a specified time period is greater than $0$
[\citet{badrwood2003}; \citet{maha2004}]. CD4 count is also dichotomized
($<$200~cells$/$mm$^2$) for each observation based on the absolute value at a
given time point or the change over a specified period. These
investigations generally include reporting of sensitivity, specificity,
positive predictive value (PPV) and negative predictive value (NPV),
where sensitivity and specificity are defined in the usual manner as
the proportions respectively of those predicted positive among those
truly positive and those predicted negative among those truly negative.
Through consideration of multiple cutoff points for both predictor and
outcomes, ROC curves are generated that illustrate the trade-off
between sensitivity and specificity.\looseness=1

Logistic regression models have also been described as a useful tool in
this setting [\citet{bagc2007}; \citet{spac2003}]. These methods draw strength on
the continuous nature of the potentially predictive markers, such as
TLC, while using a dichotomized version of CD4 count. Logistic models
have the advantage of offering a framework for incorporating multiple
continuous or categorical predictor variables and accounting for the
confounding and/or effect modifying role of patient specific
demographic and clinical factors. Adjusted odds ratios are reported
from these model fits. While this approach uses more information on the
available data, it involves first dichotomizing CD4 counts and does not
include reporting of sensitivity and specificity, two clinically
appealing and relevant concepts.

An extensive literature also exists on methodologies for ROC curves as
summarized in \citet{zhoumccl2002} and \citet{pepe2000}. Within this body
of research, methods for incorporating ordinal and continuous
predictors have been described [\citeauthor{pepe1998} (\citeyear{pepe1998}, \citeyear{pepe2005}); \citet{tostbegg1988}]
as well as approaches to handling repeated marker data \citep
{emir1998}. To our knowledge, however, these methods are developed
primarily for a dichotomous outcome such as ``diseased'' or ``not
diseased.'' In our setting, both the predictor variables and outcome of
interest are continuous biomarkers, which serve as a primary motivation
for the linear mixed effects modeling approach we describe.
Specifically, we aim to incorporate and draw strength from the complete
observed response data (rather than a dichotomized version) to arrive
at a prediction rule.

Similar to our approach, methods for time-dependent ROC curves, as
described in \citet{HeagLumlPepetime2000}, aim to characterize a
time-varying clinical measure of disease progression within a
prediction framework. \citet{HeagLumlPepetime2000} provide an eloquent
approach for the setting of a survival outcome, in which the binary
indicator for disease status is potentially censored and can vary over
time, and which involves direct modeling of the sensitivity and
specificity. In our setting, the outcome of interest is a continuous
biomarker and, thus, direct modeling of the sensitivity and specificity
in this fashion is not tenable. Instead, we consider two approaches,
one that involves direct modeling of the probability that the outcome
is above a threshold and the second that approximates the prediction
rule through use of a corresponding prediction interval. Further
extensions involving modeling of time to CD4 count below a meaningful
threshold would be interesting.

Methods involving generalized linear models and mixed effects models
have been described for estimating ROC curves
[\citet{albe2007}; \citet{gats1995}; \citet{pepe22000}]. As noted by \citet{dodd2003},
PBC
in its original formulation is an approach to estimation of the area
under the ROC curve given by the probability that the response in the
group is greater than the response is another group. The setting
described herein differs, however, since here estimation is described
for the probability that an observation is greater than a given
threshold and not for the comparison of two groups. A ROC curve is then
generated based on a prediction rule that incorporates this estimated
probability. Finally, we note that our algorithm involves generating a
single ROC curve based on a set of predictors determined in a model
fitting framework. This distinguishes our strategy from approaches that
aim to identify the most predictive set of markers by evaluating the
areas under the curve across several sets of predictors, such as \citet
{biss2008}.

PBC may be a clinically useful tool for predicting whether an
individual's CD4 count will be greater than a given threshold based on
less-expensive laboratory measures, including WBC and lymphocyte
percent. For the data example presented, using the continuous range of
the CD4 data and application of the linear mixed effects model appears
to offer better predictive performance than a first stage
dichotomization and application of the generalized linear mixed model.
This is evidenced in both the resubstitution and test sample estimates
of predictive performance. For example, for a CD4 threshold of 350 and
a test sample FP rate of $4\%$, the GLMM approach results in test
sample $\mathrm{Sensitivity}=0.50$, $\mathrm{PPV}=0.78$ and $\mathrm{NPV}=0.88$. The LMM approach,
on the other hand, yields test sample $\mathrm{Sensitivity}=0.64$, $\mathrm{PPV}=0.82$
and $\mathrm{NPV}=0.91$ for the same cutoff and test sample FP rate. While we
have not demonstrated a statistically significant difference between
the two approaches, a clear trend is observed across all rules for both
the test and learning sample data.

The primary advantages of this strategy over the tools described in
Section~\ref{sec:intro} for this data setting are as follows: (1) it
allows us to draw strength from the full range of continuous outcome
data (through linear modeling) while providing us with clinically
relevant measures, such as positive predictive value (through
subsequent classification based on probability thresholds) and (2) it
allows for simultaneous consideration of unevenly spaced biomarker
measurements over time. In the example described for predicting
absolute CD4 count based on a $200$-level threshold, a positive
predictive value of $0.98$ is observed with a false positive rate of
$0.05$, suggesting this approach may be useful in developing
alternative clinical management strategies. The relatively low NPV of
$0.36$ suggests that the approach described herein may serve best as a
prioritization tool that allows for the reduction in higher-end
capacity testing, while not replacing the use of these
tests.\looseness=1

The clinical utility of this tool, however, will require further
consideration of additional clinical and environmental factors as well
as an in-depth analysis of a diverse array of cohorts. For example, the
application presented in Section~\ref{sec:example} is based on data
from the London cohort in which a median baseline CD4 count of $219.5$
is observed. Baseline CD4 counts at initiation of therapy tend to be
lower in resource poor settings since treatment guidelines in these
settings impose a lower threshold for starting ARTs. The implication of
differing patient level characteristics such as baseline CD4 count on
the appropriateness of this approach as a diagnostic tool still
requires thorough assessment. Stratified analyses may also be
informative in identifying subgroups for which the tool is best suited.
For example, characterizing the relative performance among viremic and
nonviremic patients, or during earlier and later exposure to ARTs, will
provide additional insight into the large-scale relevance of this
approach. In addition, the example presents a prediction for each
observation within an individual. Characterizing this approach for
predicting that any of an array of observations for an individual will
be above the threshold would provide further insight into its utility.
Finally, it may be useful to additionally incorporate the acquired CD4
counts of those individuals who are tested because they are predicted
to be below the threshold. We are currently investigating these
alternative questions and settings.

The PBC approach we describe relies heavily on observing baseline CD4
counts. We are currently exploring application of this approach to data
arising from the Women's Interagency HIV Study (WIHS) and the
Multicenter AIDS Cohort study (MACS) cohorts in which dates of
initiation of therapy are observed only within a six month window. This
presents an additional challenge since our model includes a rapid rise
in CD4 counts over the first one month of therapy followed by a slower
sustained increase. Thus, in its current formulation, the precise time
of ART initiation is crucial. Further extensions may provide tools
necessary for these alternative settings; however, collection of
baseline CD4 count data at initiation of therapy for HIV is routine in
most settings and, thus, this does not diminish the potential relevance
of PBC for this application.

We also note that the proposed PBC framework is not limited to the
choice of design matrices given in Section~\ref{sec:example}.
Incorporation of additional potentially clinically relevant variables
such as sex and weight in the model fitting stage is straightforward.
As the model fit improves and the prediction variance decreases, the
value of $\alpha$ in equation~(\ref{eq:predrule}) corresponding to the
best prediction rule will likely change. In the extreme case that the
prediction variance tends to $0$, we have that $l_{ij,\alpha}$ of
equation~(\ref{eq:predrule2}) approaches $\widehat{y}_{ij}$ regardless of
$\alpha$. In this case, since the observed and predicted values would
be very close, all prediction rules would perform equally well with
sensitivity and specificity close to unity. In addition, alternative
more sophisticated models may offer improved accuracy. For example,
\citet{chu2005} describe a Bayesian random change point model for
predicting CD4 trajectories that includes both population and
individual level change points. Incorporating this modeling approach
into the PBC framework introduces the additional analytic challenge of
predicting individual-level change points for new patients and is a
direction of potential future development.

In summary, through combining modeling and an ROC curve approach, PBC
provides a flexible statistical framework for appropriately modeling
continuous biomarker data using all available data on the biomarker as
well as additional, potentially relevant continuous or categorical
predictors. At the same time, it offers interpretable measures of
diagnostic accuracy based on clinically determined thresholds. Notably,
improved prediction of CD4 count based on less-expensive and more
widely available laboratory measures, such as lymphocyte percentage and
white blood cell count, may have broad public health implications. A
sound diagnostic tool could provide for more targeted CD4 testing
strategies, offering a much needed instrument in a resource limited
setting where HIV/AIDS presents the greatest burden.

\printaddresses

\end{document}